# Dynamic Modulation of the Transport Properties of the LaAlO$_3$/SrTiO$_3$ Interface Using Uniaxial Strain

Fan Zhang[1+], Yue-Wen Fang[2+], Ngai Yui Chan[1], Wing Chong Lo[1], Dan Feng Li[1], Chun-Gang Duan[2], Feng Ding[3], Ji Yan Dai[1]*

[1]Department of Applied Physics, The Hong Kong Polytechnic University, Hung Hom, Kowloon, Hong Kong, P. R. China

[2]Key Laboratory of Polar Materials and Devices, Ministry of Education, East China Normal University, Shanghai 200062, P. R. China

[3]Institute of Textiles & Clothing, The Hong Kong Polytechnic University, Hung Hom, Kowloon, Hong Kong, P. R. China

[+] These authors contributed equally to this work.

* E-mail: jiyan.dai@polyu.edu.hk

ABSTRACT

Among the interfacial transport modulations to the LaAlO$_3$/SrTiO$_3$ (LAO/STO) heterostructure, mechanical strain has been proven to be an effective approach by growing the LAO/STO films on different substrates with variant lattice mismatches to STO. However, this lattice-mismatch induced strain effect is static and biaxial, hindering the study of strain effect in a dynamic way. In this work, for the first time, we realize dynamic and uniaxial strain to the LAO/STO oxide heterostructure at low temperature, through mechanical coupling from a magnetostrictive template. This anisotropic strain results in symmetry breaking at the interface, and induces further splitting of electronic band structure and therefore produces different conductivities along the *x*- and *y*- in-plane directions. Particularly, we observe that along the strained direction the interface conductivity decreases up to 70% under a tensile strain, while it increases 6.8% under a compressive strain at 2 K. Also, it is revealed that the modulation on the interfacial transport property can be anisotropic, i.e., the resistance changes differently when an excitation current is parallel or perpendicular to the strain direction. This approach of strain engineering provides another freedom of control for transport properties of oxide heterostructures, and opens a new way to investigating strain effect in material science.

# I. INTRODUCTION

Since the discovery of the two-dimensional electron gas (2DEG) at the $LaAlO_3/SrTiO_3$ (LAO/STO) heterointerface, there has been tremendous attractions for researchers to investigate modulation of transport properties of this heterostructure from different aspects[1-7]. Remarkable properties like giant Seebeck coefficient, superconductivity, photovoltaic effect and magnetic properties, etc., have been reported[8-15]. Among these discoveries, the effect of polar molecules on the surface of the LAO/STO and enhanced photo-response have been shown to be representative examples for possible device applications[16-18]. Among the interfacial transport modulations to the LAO/STO heterostructure, mechanical strain modulation on the 2DEG has attracted great deal of attention. Both theoretical considerations and experimental studies have been carried out, however, revealing conflict observations with the enhancement or suppression of the interfacial conductivity when a strain is applied[19-23]. Experimental efforts have been made by growing LAO/STO interfaces on different perovskite substrates to constrain the in-plane lattice constants of STO and LAO to various values. However, during the heteroepitaxial growth, the quality and surface state of the grown STO may have chance to be introduced with artificial effects, such as variation of growing temperature, oxygen pressure, and therefore, there might be some uncertainties to the film crystallinity and other microstructural defects that deviate the characterization results. Furthermore, the epitaxy induced strain is essentially isotropic, manifested by simultaneous changes in both **a** and **b** crystallographic directions. This hinders the way to unambiguously study the effect of a uniaxial strain on the possible asymmetric transport properties due to even lower crystal symmetry. Also, the static epitaxial strain cannot provide a dynamic control on the interface and thus the 2DEG.

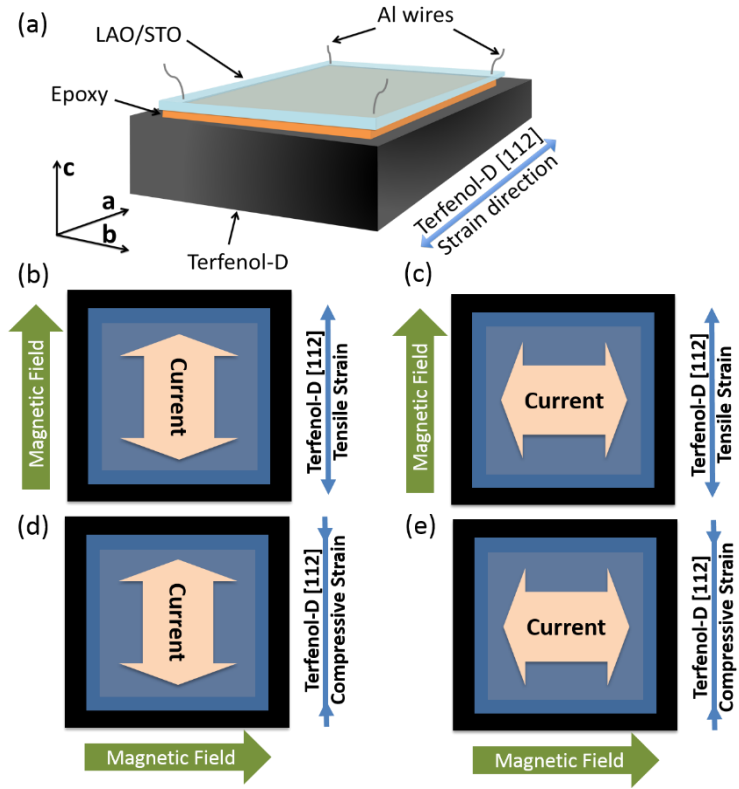

**Figure 1.** Schematic illustration of anisotropic strain generation and resistance measurement. (a) The device setup of the LAO/STO attached to Terfenol-D. (b) Measuring current is applied in parallel and (c) in perpendicular to the tensile strain. (d) Measuring current is applied in parallel and (e) in perpendicular to the compressive strain.

In this work, we report a dynamic and uniaxial modulation of strain to the LAO/STO heterostructure using a magnetostrictive material, Terfenol-D ($Tb_xDy_{1-x}Fe_2$ (x ~ 0.3)), as a template to introduce dynamic mechanical strains to the STO substrate. The device structure of LAO/STO/Terfenol-D is illustrated in Fig. 1a. The Terfenol-D is a material that can be elongated or compressed with a DC magnetic field, while the strain type and strength can be modulated by the magnitude and direction of the field[24] (details can be found in Supplemental Information[25]). Then the LAO/STO was mounted to Terfenol-D with epoxy after STO was thinned to 100 μm. By applying a magnetic field, a tensile or compressive strain can be applied to the LAO/STO heterostructure. Under different types of strains, transport properties were measured with current supplied parallel or perpendicular to the strain direction as illustrated in Figs. 1b-e. Electrodes are aluminum wires ultrasonically bonded at four corners of the sample, while the two that on the same side of current injection or extraction were shorted to adopt a bar configuration. The resistance is calculated as the ratio of a measured voltage and the supplied current. At 2 K, we observe that along the strain direction the interfacial conductivity decreases up to 70% under a tensile strain, while it increases 6.8% under a compressive strain. These results also reveal that the modulation on the interfacial transport property can be anisotropic, i.e., the resistance changes differently when the current is parallel or perpendicular to the strain direction. This approach of strain engineering provides another freedom of control for transport properties of oxide heterostructures.

## II. EXPERIMENTS AND RESULTS

### A. Sample preparation and characterization

LaAlO$_3$ thin films were grown on TiO$_2$ terminated SrTiO$_3$ (001) substrate using a RHEED-assisted laser molecular beam epitaxy (Lambda Physik COMPex 205, wavelength = 248 nm) with conditions as reported: the deposition process was conducted at 750 °C of temperature and $2\times10^{-5}$ Pa of vacuum degree, followed by an in-situ annealing at 550 °C with 1000Pa of O$_2$ for 1 hour before cooling down to room temperature[16]. After the STO substrate of LAO/STO sample being milled to ~100 μm for efficient strain transfer from the template, the LAO/STO device (2.5mm×2.5mm×0.1mm) was mounted on the magnetostrictive alloy, Terfenol-D, by M-bond epoxy which can maintain elongation capability at 2 K. Terfenol-D's magnetic easy axis [112] is aligned in-plane with the interface of LAO/STO and along edges of the sample ([010] or [100] directions). Four aluminum wires were ultrasonically bonded at corners of the square sample. To conduct the anisotropic electric characterization, each two electrodes that on the same side of current injection or extraction were shorted to simulate a bar configuration. The resistance is calculated as the ratio of a measured voltage and the supplied current. In Physical Property Measurement System (PPMS, Model 6000 by Quantum Design), an in-plane magnetic field from -5000 to +5000 Oe was applied to the Terfenol-D to introduce strain, at temperatures from 300 to 2 K. Under uniaxial strains generated in both parallel and perpendicular to the current, as illustrated in Fig. 1b-e, the interfacial resistance was measured with a constant current of 500 nA. Control samples without Terfenol-D were also thinned and measured in the same manner.

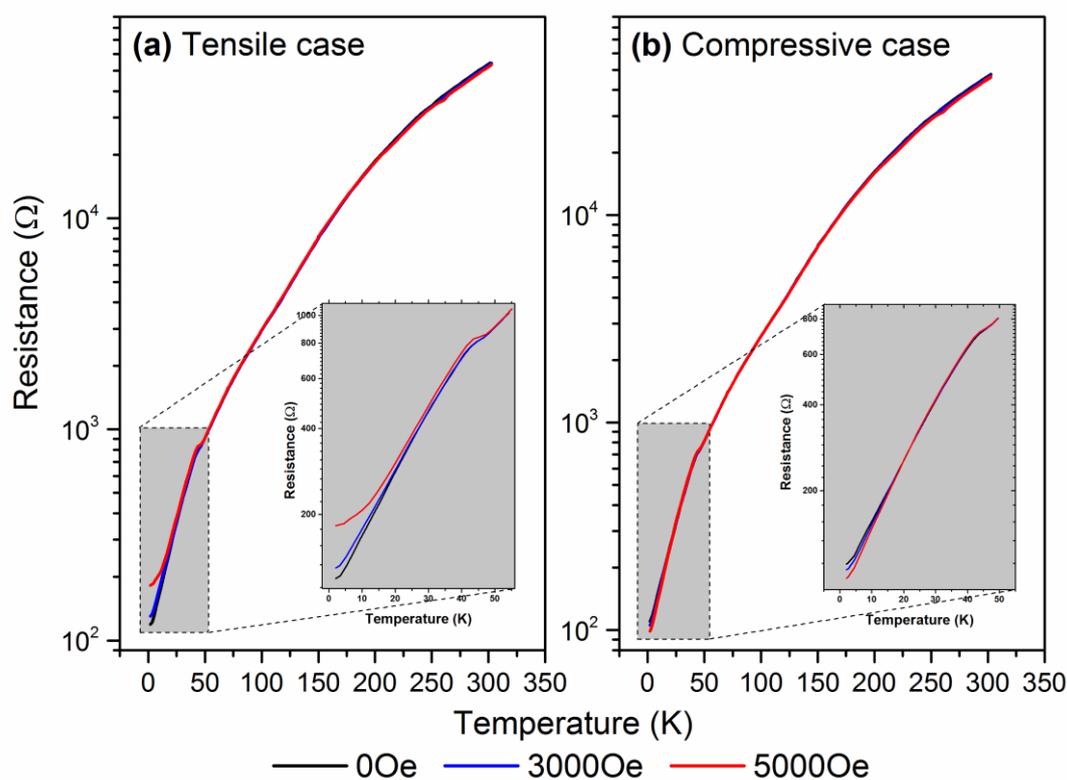

**Figure 2.** Temperature-dependent resistance (*R-T*) curves in variant magnetic field and strain types at temperatures ranging from 2K to 300K. (a) Tensely strained LAO/STO (b) Compressively strained LAO/STO.

Under different strains that driven by variant magnetic field, we measured the temperature-dependent resistance (*R-T*) curve. From Fig. 2, it can be seen that when a strain is applied to the LAO/STO heterostructure through magnetostrictive effect, with a current being parallel to the uniaxial strain direction, the conductivity of the interfacial 2DEG shows obvious change, especially when temperature is lower than 50 K. As illustrated in Fig. 2a, with a tensile strain driven by magnetic field of 0, 3000 and 5000 Oe, respectively, the difference in the *R-T* curve emerges below 50 K. For the case of a compressive strain, the *R-T* curve shown in Fig. 2b starts splitting below 30 K. At the same time, the relative resistance change at high temperatures is negligible. We note that especially in the low temperature regime, the conductivity of 2DEG is significantly tailored with a dynamic strain. At higher temperatures, the strain also shows influence on the transport properties, but much smaller in contrast to that at low temperatures. At room temperature the resistance is high, so the relative change of *R* is very small. The phenomenon that there is a sharp increase of resistance of LAO/STO at high temperatures could be explained as a consequence of complex coupling of lattice, charge and phonon[26].

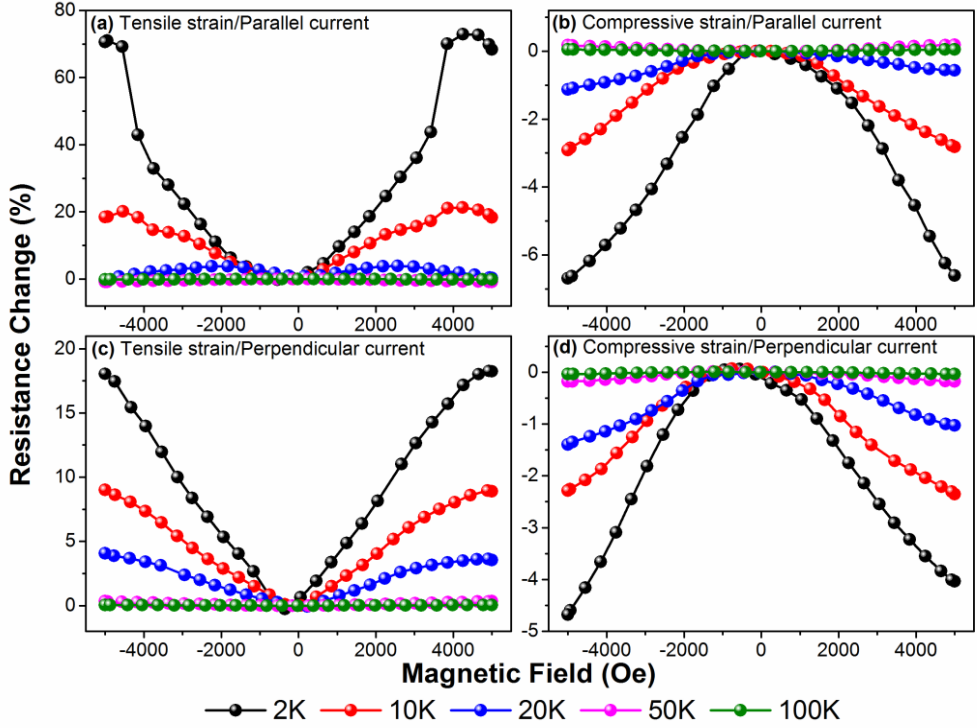

**Figure 3.** Resistance change $(R-R_0)/R_0$ in percentage of a LAO/STO/Terfenol-D sample. (a) Tensile strain being parallel to current, as Fig. 1b shows; (b) Compressive strain being parallel to current, as Fig. 1d shows; (c) Tensile strain being perpendicular to current, as Fig. 1c shows; (d) Compressive strain being perpendicular to current, as Fig. 1e shows. ($R_0$ is the resistance at the unstrained state)

At different temperatures, the resistance of the interface was also measured with a sweeping magnetic field from -5000 to +5000 Oe, a range where Terfenol-D can be sufficiently elongated. Fig. 3 shows the resistance as a function of the magnetic field (i.e. strain) of the sample for different temperatures from 2 to 100 K. We observe that the resistance appears to be very sensitive to the applied strain, and the resistance modulation is particularly obvious at low temperatures, which can also be seen in Fig. 2. In Fig. 3a, one can see that the resistance increases by 70 % at 2 K when the sample is under a tensile strain. By contrast, the resistance decreases 6.8% under a compressive strain, as shown in Fig. 3b.

Upon these observations, we systematically investigated the resistance modulations with different directions of measuring current, which can be applied either parallel or perpendicular to the strain direction. As it can be seen from Figs. 3a-d, the behaviors of the change in resistance are different and strongly depend on the strain type and current direction. Figs. 3a and 3b show the magnetic field strength-resistance measured with current applied in the **a** direction which is parallel to the strain; while Figs. 3c and 3d show the magnetic field strength-resistance measured with current applied in the **b** direction which is perpendicular to the strain. It is worth noting that, with different current directions, there is no signature of sign inversion of resistance change when the same type of strain presents. However, at the same temperature and strain state, the relative resistance change is smaller for the case with the current applied perpendicularly to the strain direction. At higher temperatures (200 and 300 K), the changes in resistance are very small (< 0.1%) (as shown in Fig. S1 in Supplemental Information[27]). In previous reports[28-30], very small magnetoresistance were observed at the same configuration of current and field. Therefore, at temperatures above 100 K, it is difficult to distinguish the strain effect from a pure magneto-resistance contribution. Control samples also show magnetoresistance (MR) effect, but it is less than 0.5% when current is parallel to magnetic field, and less than 1.5% when current is perpendicular to magnetic field (as shown in Fig. S2 in Supplemental Information[31]). The MR results are consistent with previous reports[28-30].

B. Theoretical moldeling

Based on these findings, we claim that a dynamic modulation on transport properties of the 2DEG has been realized with a strain from the magnetostrictive coupling. The general trend of change in resistance with a tensile strain is similar to that of the reported case where LAO was grown on tensely strained STO[19,20]. While for the compressive strain case, our result is different, where we observed a decrease of resistance. To elucidate the uniaxial strain effect on the transport properties of LAO/STO in theory, which has never been reported before, first-principles calculations within the framework of density of functional (DFT) method implemented in Vienna *ab initio* Simulation Package (VASP)[32,33] were carried out. In calculations, we use local density approximation+Hubbard $U$ (LDA+$U$) method, where $U = 5$ eV for Ti-*d* orbitals and $U = 11$ eV for La-*f* orbitals are adopted to describe the strongly correlated electronic states. The kinetic energy cutoff is set to be 500 eV. Brillouin zone integration is sampled using Γ centered Monkhorst-Pack grid of $10 \times 10 \times 1$ k points in self-consistent field calculations for LaAlO$_3$/SrTiO$_3$ models. For the unstrained LaAlO$_3$/SrTiO$_3$ slabs, the in-plane lattice constants are fixed at the optimized lattice constant of bulk SrTiO$_3$ (3.904 Å obtained by LDA+$U$). All coordinates of atomic positions along the **c**-direction perpendicular to the interface are fully relaxed until the forces are less than 0.01 eV/ Å and concurrently the energy convergence criterion 10E−6 eV is ensured. In particular, the effects on electronic properties brought by the in-plane compressive/tensile strains at absolute zero temperature were investigated. In order to obtain obvious trend of strain-induced resistance changes, we applied 1% and 3% tensile and compressive strains in the calculations (even though the Terfenol-D can only induce a strain less than 1%). For comparison, unstrained structure was also considered. We employ a symmetrical (LaAlO$_3$)$_4$/(SrTiO$_3$)$_{7.5}$/(LaAlO$_3$)$_4$ model with a vacuum of 20 Å to separate the periodically repeated slabs. In the model, the strain is applied to lattice constant a along *x* direction, meanwhile lattice constant b along *y* direction is varied linearly according to the Poisson's ratio[34].

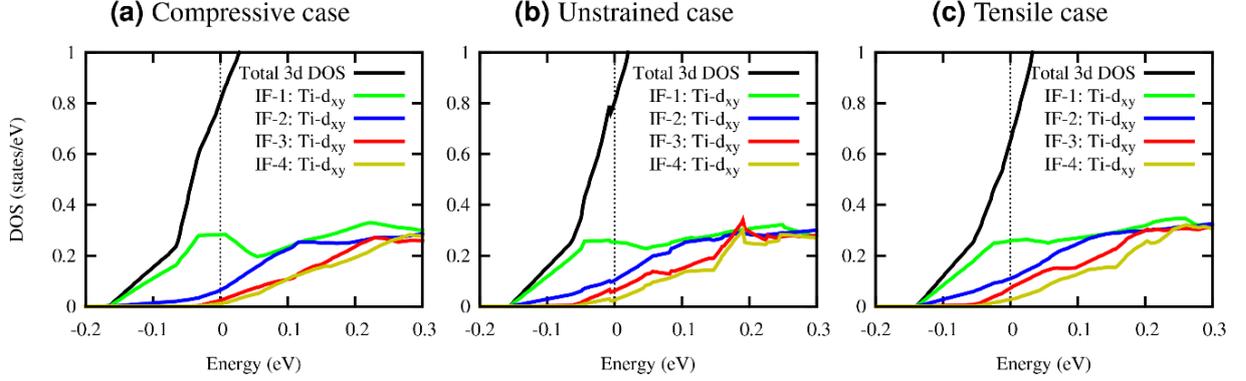

**Figure 4.** Ti-$d_{xy}$ orbital resolved partial density of sates and total density of states (DOS) of Ti-$3d$ in each case. (a) 3% compressive strain. (b) Unstrained strain. (c) 3% tensile strain. Zero in energy axis is the reference for the Fermi level. IF-$N$ ($N$ = 1, 2, 3, 4) is the index of STO interfacial layer. Total $3d$ DOS shown here is the summation of all the Ti-$3d$ DOS from the four STO interfacial layers.

The obtained density of states (DOS) is shown in Fig. 4, where zero is the reference for Fermi level. The total $3d$ DOS of Ti atoms show that Fermi level is shifted toward the conduction bands in compressively strained LAO/STO with (Fig. 4a) respect to that in the unstrained LAO/STO (Fig. 4b). Nevertheless, it is shifted toward valence bands in the tensile case (Fig. 4c), i.e., 3% compressive strain raises the charge carrier population and 3% tensile strain reduces it. In order to demonstrate this result quantitatively, one can estimate the carrier charge concentration by integrating the DOS around the Fermi level for each case. The estimated carrier concentrations for compressive, unstrained and tensile cases are $3.14 \times 10^{13}$ cm$^{-2}$, $2.87 \times 10^{13}$ cm$^{-2}$ and $2.20 \times 10^{13}$ cm$^{-2}$, respectively, indicating that the uniaxial strains can indeed effectively modulate the interfacial conductance. Besides, it can be found that Ti-$d_{xy}$ states with strong two dimensional characteristics dominate the interfacial electronic sates in all cases from the Ti-$d_{xy}$ orbital resolved partial density of sates (PDOS) of in Fig. 4. The shift of band minima of PDOS shows a downward band bending at the STO side, being suggested to be the key of formation of 2DEG at LAO/STO interfaces[35]. Especially in compressive case, the quantum well created by the band bending is deepest, which can confine more electrons at the interface; by contrast, less electrons can be trapped in the tensile state because of the narrower quantum well compared to that in unstrained LAO/STO.

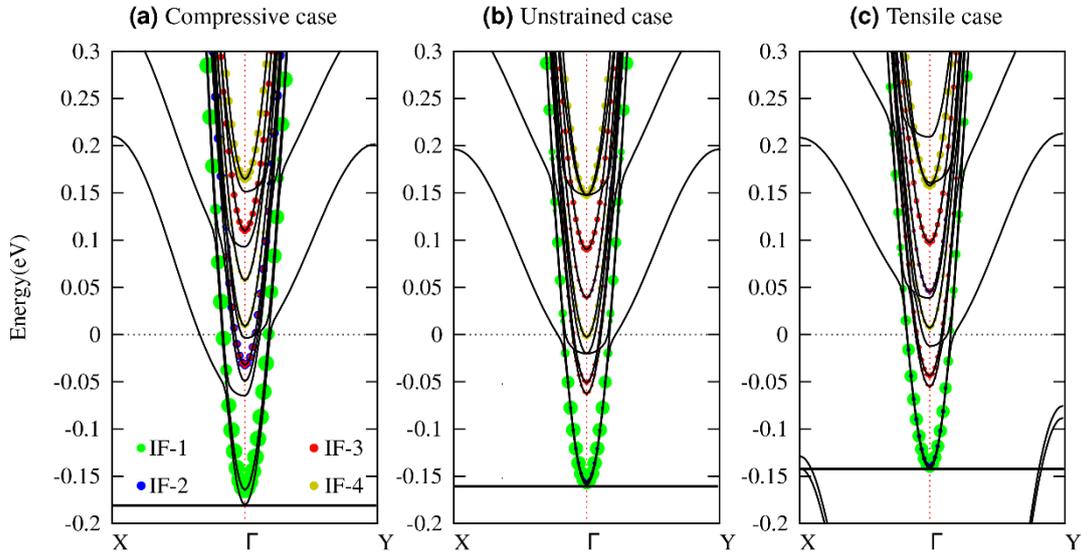

**Figure 5.** Ti-$d_{xy}$ weighted electronic band structures in each case. (a) 3% compressive strain. (b) Unstrained strain. (c) 3% tensile strain. Zero in energy axis is the reference for the Fermi level. The colors of solid circles denote $d_{xy}$ contributions from different STO layers.

Apart from the changes of interfacial conductance seen in Fig. 4, another experimentally observed phenomenon, that the anisotropic interfacial conductance is introduced by the uniaxial strains, can also be revealed by the first-principles calculations. Fig. 5 shows the band structures weighted on Ti-$d_{xy}$ orbitals obtained in calculations. In unstrained LAO/STO, the preserved C4v symmetry results in a symmetrical band structure (Fig. 5b). On the other hand, asymmetrical band structures are induced (Figs. 5a and 5c) owing to the reduction of symmetry by applying uniaxial strains. This symmetry breaking drives LAO/STO into novel 2DEG systems with more complex subband structures at the interfaces. Consequently, electron mobility, which is inversely proportional to the effective mass, can be significantly different in *x* and *y* directions when uniaxial strain is applied. We thus calculate the effective masses $m_x^*$ and $m_y^*$, as plotted in Fig. 6. When it is in a free state (0%), $m_x^*$ strictly equals $m_y^*$, indicating isotropic in-plane conductance. However, after applying uniaxial strains, notable difference of $m_x^*$ and $m_y^*$ can be clearly observed, reflecting the emergence of anisotropic interfacial conductance. In particular, with the increase of compressive strain along *x* direction, the effective mass of electrons tends to be lighter in the *x* direction and it is heavier in the *y* direction, respectively. However, the effective masses in the cases with tensile strains are precisely the opposite. In addition, it can also be seen from Fig. 6 that the electron mobility is much more sensitive to tensile strains than that to compressive strains. These results qualitatively explain the large difference for anisotropic resistance change (maximumly 70% and 18% as shown in Figs. 3a and 3c) under tensile strains and smaller difference for anisotropic resistance change (maximumly 6.8% and 4.7% as shown in Figs. 3b and 3d) under compressive strains at 2 K.

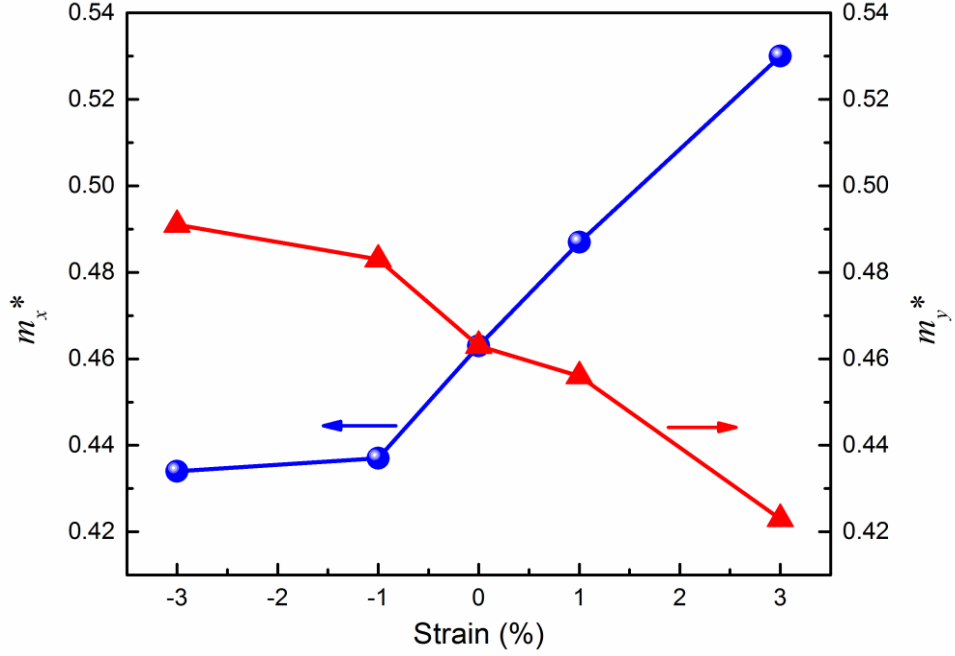

**Figure 6.** Effective masses $m_x^*$ (blue point) and $m_y^*$ (red triangle) in the parallel ($x$) and transverse ($y$) directions. "-" and "+" denote "compressive strain" and "tensile strain", respectively.

## III. DISCUSSIONS

To present an evidence that the strain is truly transferred to STO and modulates the 2DEG, we grinded a bare (001) STO substrate into 30 um and made it into a parallel plate capacitor with Au electrode at both sides, and this capacitor was attached on Terfenol-D as a strain sensor and control sample. As expected, a tensile strain enlarges the capacitor's area and decreases the thickness and therefore increases the capacitance, while a compressive strain induces a capacitance decrease. By the inversed capacitance change, we proved the presence of tensile and compressive strain on STO substrate (more details are provided in the part 4 of Supplemental Information[36]). It should be noticed, however, that it is very difficult to quantitatively measure the strain that being transferred to the STO substrate especially at temperatures as low as 2 K, where most strain gauges do not work.

One may argue that the simulation results only show a small change of resistance under strains. This large mismatch can be further understood by considering the generation of polarization in STO substrate under strain. Since the compressive strain-induced resistance change is very small (the compressive strain is much smaller for Terfenol-D when a perpendicular magnetic field to [112] direction is applied), here we only consider the tensile strain case which induces a very large increase of resistance. As claimed in the report where LAO was grown on epitaxialy strained STO[19], the tensile strain destroys the 2DEG. Biegalski's work reported that a STO layer under tensile strain on $DyScO_3$ substrate behaves as a relaxor with presence of nanoscale polar regions[37]. We believe that such nanoscale polar regions could form local field and potential barriers throughout the interface and localize the conducting interfacial electrons by scattering. This will contribute more to the increase of the interfacial resistance. Therefore, when considering the strain modulation to the interfacial conductivity at the LAO/STO interface, the STO polarization effect under strain should also be considered.

## IV. CONCLUSION

To conclude, we demonstrate a dynamic and uniaxial strain induced modulation to the transport properties of the 2DEG at the conducting LAO/STO interface. An increase in the interfacial resistance has been observed when a uniaxial tensile strain is applied to the LAO/STO interface, while a uniaxial compressive strain produces a resistance decrease. The conductivity changes anisotropically when the measuring current is applied parallel or perpendicular to the strain direction. These results reveal the fact that appropriate anisotropic strains can result in anisotropic electron mobility and effective mass along different directions. This study provides further insights into the LAO/STO systems and paves the pathway to designing novel devices made of complex oxide interfaces possessing more exotic properties. Furthermore, this method can also be applied to ultrathin films and two dimensional materials for a dynamic and uniaxial strain effect study, especially at cryogenic temperatures. We hope this method can motivate more discoveries not only on LAO/STO systems, but also stimulate new and striking findings in low dimensional materials in other fields such as the field of spintronics and superconductivity.

ACKNOWLEDGMENT

This work was supported by the NSFC/RGC grant (N-PolyU517/14), and The Hong Kong Polytechnic University Strategic Importance Project (1-ZE25). The work at ECNU was supported by the National Basic Research Program of China (2014CB921104) and NSFC (61125403). All computations were conducted in computing center at ECNU and Chinese Tianhe-1A system at the National Supercomputer Center. The Terfenol-D alloys are provided by Dr. CHOY Siu Hong.

# Supplemental Information

Fan Zhang[1+], Yue-Wen Fang[2+], Ngai Yui Chan[1], Wing Chong Lo[1], Dan Feng Li[1], Chun-Gang Duan[2], Feng Ding[3], Ji Yan Dai[1]*

[1]Department of Applied Physics, The Hong Kong Polytechnic University, Hung Hom, Kowloon, Hong Kong, P. R. China

[2]Key Laboratory of Polar Materials and Devices, Ministry of Education, East China Normal University, Shanghai 200062, P. R. China

[3]Institute of Textiles & Clothing, The Hong Kong Polytechnic University, Hung Hom, Kowloon, Hong Kong, P. R. China

[+] These authors contributed equally to this work.

* E-mail: jiyan.dai@polyu.edu.hk

1. **Spontaneous magnetostriction of Terfenol-D**

   Terfenol-D ($Tb_xDy_{1-x}Fe_2$ (x ~ 0.3)) is a widely used magnetostrictive material since its invention. Terfenol-D alloy is formed with magnetic domains. These magnetic domains rotate towards the direction of the externally applied low level magnetic field. With such microstructure realignment like reversible domain wall movement and momentum rotation, large macrostructure elongation or contraction can be observed[1]. By applying a magnetic field parallel or perpendicular to its [112] easy axis, as shown in Fig. 1, a tensile or compressive strain can be generated along the [112] direction. For a Terfenol-D fiber, when a parallel magnetic field is applied, it will elongate. The fiber can also contract when a perpendicular field is applied. As shown in Fig. 1 in the main text, we assume that the Terfenol-D easy axis [112] is along the **a** direction, where the magnetostriction is the largest. When the magnetic field is applied along the **a** direction, a tensile strain will be generated. Under the same mechanism, a compressive strain in the **a** direction can be observed when the magnetic domains rotate towards the **b** direction, in case of a magnetic field is applied in **b**. The strain is qualitatively believed to be proportional to the square magnitude of the applied magnetic field at relatively low field. The strain can be transferred to the LAO/STO interface through the mechanical coupling and thus the transport properties is modulated.

2. **Resistance change of the strained LAO/STO/Terfenol-D at temperatures above 100K.**

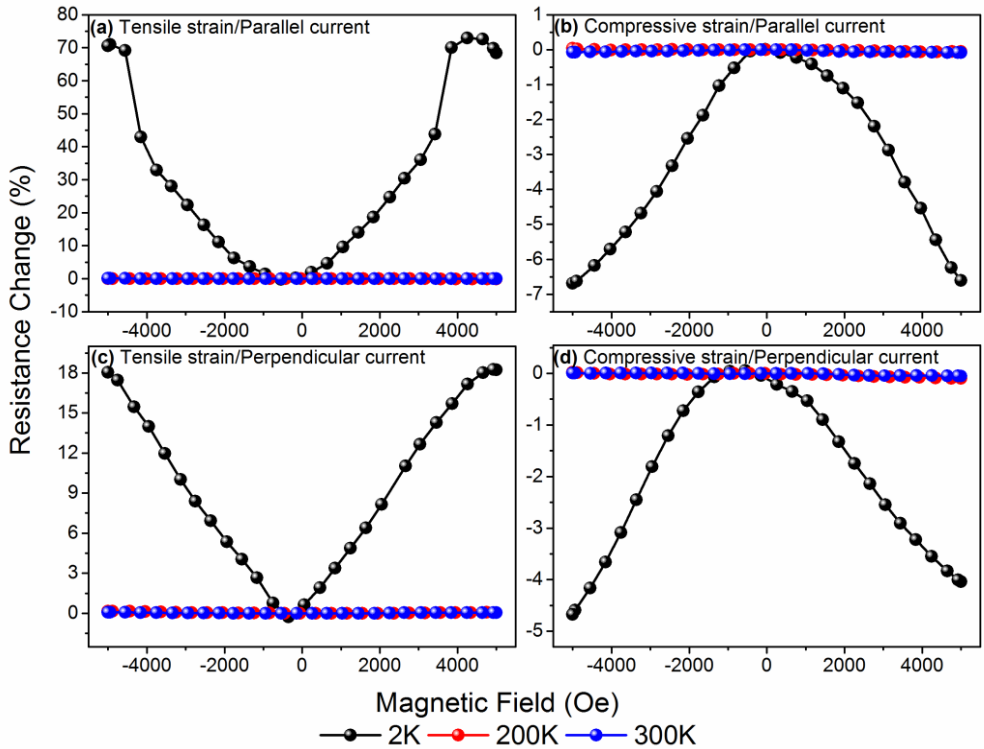

**Figure S1.** Resistance change $(R-R_0)/R_0$ in percentage of a LAO/STO/Terfenol-D sample at variant temperatures. (a) Tensile strain being parallel to excitation current, as illustrated in Fig. 1b; (b) Compressive strain being parallel to excitation current, as illustrated in Fig. 1d; (c) Tensile strain being perpendicular to excitation current as illustrated in Fig. 1c; (d) Compressive strain being perpendicular to excitation current, as illustrated in Fig. 1e.

3. **Control sample's magnetoresistance**

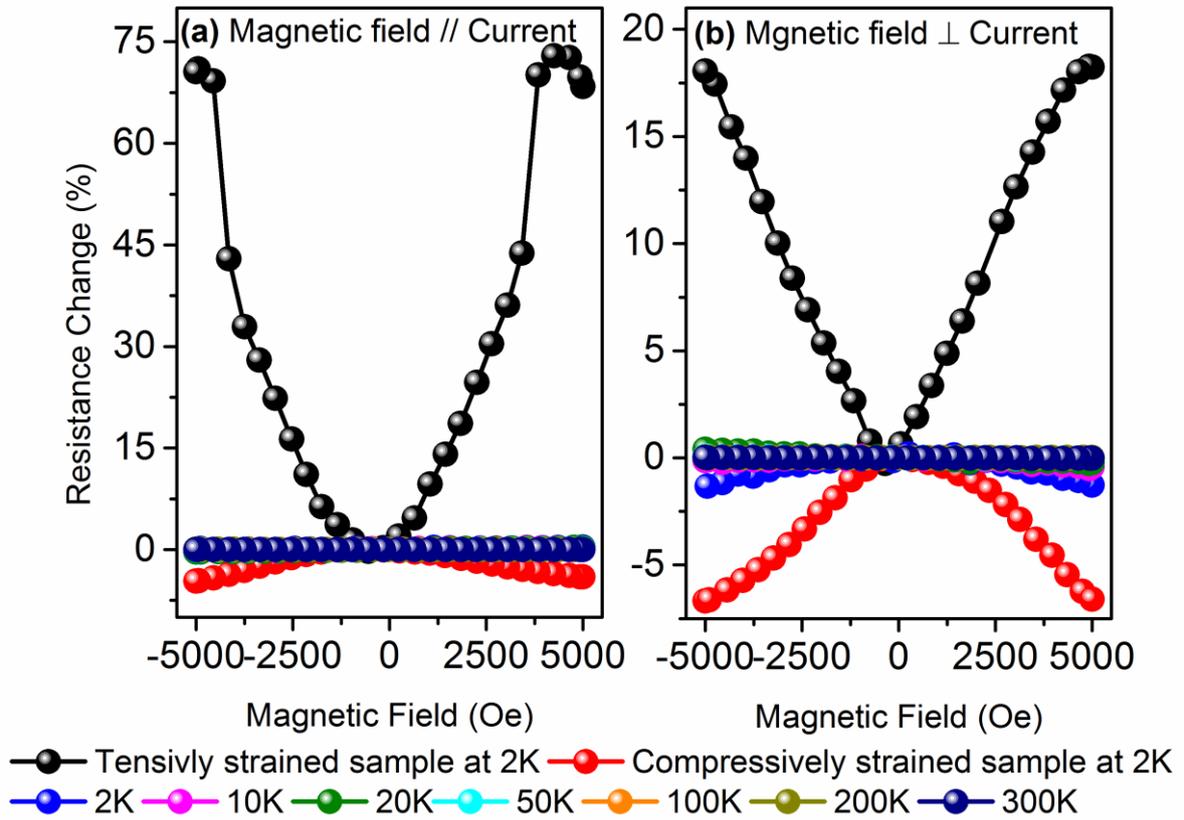

**Figure S2**. Comparison of resistance change $(R-R_0)/R_0$ in percentage of strained sample at 2K and control samples LAO/STO. (a) Magnetic field being parallel to excitation current. (b) Magnetic field being perpendicular to excitation current.

4. **An evidence of the strain is transferred to STO**

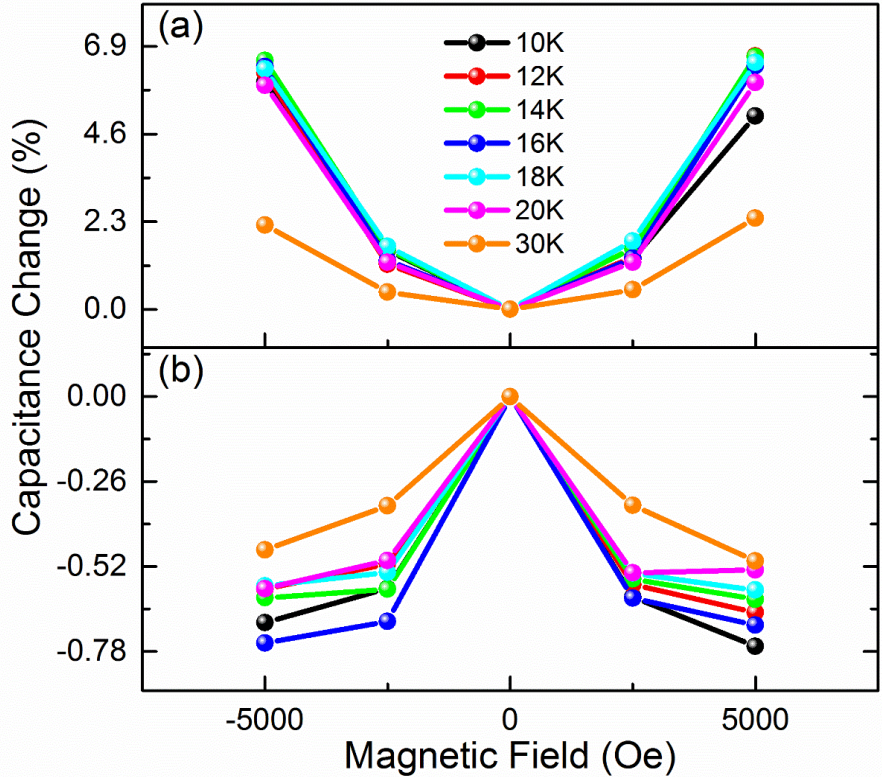

**Figure S3.** The capacitance change $(C-C_0)/C_0$ in percentage under variant Magnetic field. (a) Tensile strain case. (b) Compressive strain case.

To prove that the strain is transferred to the STO substrate, we did the following experiment. A bare STO substrate (instead of LAO/STO) is thinned and coated with Au on both side, which forms a capacitor with STO as a dielectric. With the same method described in the manuscript, the capacitor is attached to Terfenol-D, which is expected to stretch or compress the capacitor and induce a capacitance change.

The Figure S3 above shows the capacitance change when the STO is strained. As illustrated, the capacitance increases when magnetic field is generated to induce tensile strain, while a decrease trend is shown in the compressive case. The change is percentage is larger under a tensile strain.

We know that the capacitance of a capacitor constructed of two parallel plates both of area *A* and separated by a distance *d* is determined by the equation:

$$C = \varepsilon_r \varepsilon_0 \frac{A}{d}$$

where $\varepsilon_0$ is the electric constant and $\varepsilon_r$ is the relative permittivity, or dielectric constant. The equation can be adapted as: the capacitance measured is related to the dielectric constant $\varepsilon_{STO}$ and configuration of the capacitor. Let's hypothesize that there is no strain transferred to STO, so the configuration will not change under a static magnetic field. Then the capacitance change should be attributed to a change of $\varepsilon_{STO}$ under magnetic field, which should be a same situation when the magnetic field is generated along either [010] or [100] (different type). However, from the data above, a different change trend is shown under the two cases. The hypothesis is disconfirmed, which reveals that the configuration is changed because of strain. What's more, the trend of capacitance change is consistent with the expectation: under a tensile strain, the area A will increase with the distance d will decrease which can lead to a capacitance increase; for a compressive strain case, the situation is inversed. It is still hard to quantify the strain with the data above, because it is very possible that the electric constant of STO is tuned under strain, which has been found in an epitaxially strained STO[2].With the above explanation, we hope the experiment could serves as an evidence to prove that strain induced by Terfenol-D is truly transferred to STO and modulates the 2DEG. The capacitance as a function of magnetic field was measured in PPMS equipped with an HP4294A impedance analyzer.